\documentstyle[aps,eqsecnum,preprint,floats,epsf,epsfig]{revtex}
\begin{document}
\def\be{\begin{eqnarray}}
\def\en{\end{eqnarray}}
\def\non{\nonumber}
\def\la{\langle}
\def\ra{\rangle}
\def\nc{N_c^{\rm eff}}
\def\vp{\varepsilon}
\def\vma{{_{V-A}}}
\def\vpa{{_{V+A}}}
\def\m{\hat{m}}
\def\ov{\overline}
\def\pr{{\sl Phys. Rev.}~}
\def\prl{{\sl Phys. Rev. Lett.}~}
\def\pl{{\sl Phys. Lett.}~}
\def\np{{\sl Nucl. Phys.}~}
\def\zp{{\sl Z. Phys.}~}
\def\lsim{ {\ \lower-1.2pt\vbox{\hbox{\rlap{$<$}\lower5pt\vbox{\hbox{$\sim$}
}}}\ } }
\def\gsim{ {\ \lower-1.2pt\vbox{\hbox{\rlap{$>$}\lower5pt\vbox{\hbox{$\sim$}
}}}\ } }

\font\el=cmbx10 scaled \magstep2 {\obeylines \hfill August, 1999}

\vskip 1.5 cm

\begin{center}
{\large {\bf The $\Delta I=1/2$ Rule in Kaon Decays: }}\par
{\large {\bf A New Look}}
\vskip 1.0cm {\bf Hai-Yang Cheng}
\vskip 0.5cm Institute of Physics, Academia Sinica,\par
Taipei, Taiwan 105, Republic of China
\end{center}
\vskip 1.0cm

%PACS numbers: 12.38.Cy, 11.10.Hi

\centerline{\bf Abstract} \vskip 0.3cm {\small The $K\to\pi\pi$
decay amplitudes are studied within the framework of generalized
factorization in which the effective Wilson coefficients are
gauge-invariant, renormalization-scale and -scheme independent
while factorization is applied to the tree-level hadronic matrix
elements.  Nonfactorized contributions to the hadronic matrix
elements of $(V-A)(V-A)$ four-quark operators, which are needed to
account for the suppression of the $\Delta I=3/2$ $K\to\pi\pi$
amplitude $A_2$ and the enhancement of the $\Delta I=1/2$ $A_0$
amplitude, are phenomenologically extracted from the measured
$K^+\to\pi^+\pi^0$ decay and found to be large. The $A_0/A_2$
ratio is predicted to lie in the range 15-17 for $m_s(1\,{\rm
GeV})=(127-150)$ MeV. Vertex and penguin-type radiative
corrections to the matrix elements of four-quark operators and
nonfactorized effects due to soft-gluon exchange account for the
bulk of the $\Delta I=1/2$ rule. Comparison of the present
analysis with the chiral-loop approach is given.

}

\pagebreak

%%%%%%%%%%%%%%%%%%%%%%%%%%%%%%%%%%%%%%%%%%%%%%%%%%%%%%%%%
%%%%%%%%%%%%%%%%%%%%%%%%%%%%%%%%%%%%%%%%%%%%%%%%%%%%%%%%%%%%%%%

\section{Introduction}
The effective Hamiltonian approach is the standard starting point
for describing the nonleptonic weak decays of hadrons. In this
approach, the decay amplitude has the form $A\sim \sum c_i(\mu)\la
O_i(\mu)\ra$, where the renormalization scale $\mu$ separates the
short-distance contributions contained in the Wilson coefficient
functions $c_i(\mu)$ and the long-distance contributions contained
in the hadronic matrix elements of 4-quark operators $\la
O_i(\mu)\ra$. Of course, the physical amplitude should be
independent of the choice of the renormalization scale and scheme.
This means that the matrix elements have to be evaluated in the
same renormalization scheme as that for Wilson coefficients and
renormalized at the same scale $\mu$. In principle, the scale
$\mu$ can be arbitrary as long as it is large enough to allow for
a perturbative calculation of Wilson coefficients. In practice, it
is more convenient to choose $\mu$ to be the scale of the hadron
mass of the decaying particle so that the logarithmic term in the
matrix element $\la O(\mu)\ra$, which is of order $\ln(M^2/\mu^2)$
with $M$ being the hadron mass, is as small as possible, leaving
the large logarithms,  which are summed to all orders via the
renormalization group technique, to $c(\mu)$.

Since the hadronic matrix elements are very difficult to
calculate, it is not surprising that the issue of their $\mu$
dependence is generally not addressed in the literature. For meson
decays, a popular approach is to evaluate the matrix elements
under the factorization hypothesis so that $\la O(\mu)\ra$ is
factorized into the product of two matrix elements of single
currents, governed by decay constants and form factors. However,
the information of the scale dependence of $\la O(\mu)\ra$ is lost
in the factorization approximation because the vector or
axial-vector current is partially conserved and hence scale
independent. Consequently, the $\mu$ dependence of Wilson
coefficients does not get compensation from the matrix elements.
Although the correct $\mu$ dependence of $\la O(\mu)\ra$ should be
restored by the nonfactorized contributions to hadronic matrix
elements, the difficulty is that nonfactorized effects are not
amenable owing to their nonperturbative nature. Hence, the
question is can we apply factorization to the matrix elements and
in the meantime avoid the scale problem with $\la O(\mu)\ra$ ?

Fortunately, the $\mu$ dependence of hadronic matrix elements is
calculable in perturbation theory. After extracting the $\mu$
dependence of $\la O(\mu)\ra$ and combining it with the Wilson
coefficients $c(\mu)$, we obtain renormalization-scale and -scheme
independent effective Wilson coefficients. Then the factorization
approximation can be safely applied afterwards to the matrix
elements of the operator $O_i$ at the tree level.

For the case of kaon decays such as $K\to\pi\pi$, it is obvious
that $\mu$ cannot be chosen to be of order $m_K$; instead it has
to be at the scale of 1 GeV or larger so that $c(\mu)$ can be
reliably computed. Conventionally, the $\mu$ dependence of matrix
elements involving kaons and pions are calculated by considering
chiral loop corrections to $\la O\ra$. When chiral loops are
regularized using the same dimensional regularization scheme as
that for Wilson coefficients, the $\mu$ dependence of
long-distance contributions will presumably match the scale
dependence of Wilson coefficients so that the resulting physical
amplitude is $\mu$ independent. While the scale dependence of
$K\to\pi\pi$ matrix elements can be furnished by meson loops, it
is clear that this approach based on chiral perturbation theory is
not applicable to heavy meson decays. Therefore, it is desirable
to consider the nonleptonic decays of kaons and heavy mesons
within the same framework of generalized factorization in which
the effective Wilson coefficients $c^{\rm eff}_i$ are
renormalization-scale and -scheme independent while factorization
is applied to the tree-level hadronic matrix elements. The purpose
of the present analysis is to see if our understanding of the
$\Delta I=1/2$ rule can be improved in the effective Hamiltonian
approach.

The celebrated $\Delta I=1/2$ rule in kaon decays still remains an
enigma after the first observation more than four decades ago. The
tantalizing puzzle is the problem of how to enhance the $A_0/A_2$
ratio of the $\Delta I=1/2$ to $\Delta I=3/2$ $K\to\pi\pi$
amplitudes from the outrageously small value 0.9 [see Eq.
(\ref{naive}) below] to the observed value $22.2\pm 0.1$ (for a
review of the $\Delta I=1/2$ rule, see \cite{Cheng89}). In the
approach of the effective weak Hamiltonian, the $A_0/A_2$ ratio is
at most of order 7 even after the nonfactorized soft-gluon effects
are included \cite{Cheng89}. Moreover, the $\mu$ dependence of
hadronic matrix elements is not addressed in the conventional
calculation. In the past ten years or so, most efforts are devoted
to computing the matrix elements to ${\cal O}(p^4)$ in chiral
expansion. This scenario has the advantages that chiral loops
provide the necessary scale dependence for hadronic matrix
elements and that meson loop contributions to the $A_0$ amplitude
are large enough to accommodate the data. However, it also becomes
clear that chiral loops alone cannot explain the $A_2$ amplitude
(see Sec. IV). Consequently, it is necessary to take into account
nonfactorized effects on $K^+\to\pi^+\pi^0$ in order to have an
additional suppression for the $\Delta I=3/2$ transition.

Contrary to the chiral approach, the difficulty with the $\mu$
dependence of the physical $K\to\pi\pi$ amplitude is circumvented
in the present analysis by working in the effective Hamiltonian
approach in which the effective Wilson coefficients are
gauge-invariant, renormalization-scale and -scheme independent.
This approach is not only much simpler than chiral loop
calculations but also applicable to heavy meson decays. By
extracting nonfactorized effects from $K^+\to\pi^+\pi^0$, we shall
see that nonperturbative effects due to soft-gluon exchange and
perturbative radiative corrections to four-quark operators account
for the bulk of the observed $\Delta I=1/2$ amplitude.

The present paper is organized as follows. In Sec. II we construct
scheme and scale independent effective Wilson coefficients
relevant to kaon decays. The $K\to\pi\pi$ matrix elements are
evaluated in Sec. III. Comparison of the present analysis of the
$\Delta I=1/2$ rule with the chiral loop approach is made in Sec.
IV. Sec. V is for the conclusion.

\section{Framework}
The effective Hamiltonian relevant to $K\to\pi\pi$ transition is
\be
{\cal H}_{\rm eff}(\Delta S=1) &=& {G_F\over\sqrt{2}}V_{ud}V_{us}^*
\Bigg(\sum^{10}_{i=1}c_i(\mu)O_i(\mu)\Bigg)+{\rm h.c.},
\en
where
\be
c_i(\mu)=z_i(\mu)+\tau y_i(\mu),
\en
with $\tau=-V_{td}V_{ts}^*/(V_{ud}V_{us}^*)$, and
\be  \label{Oi}
&& O_1= (\bar ud)_\vma(\bar su)_\vma, \qquad\qquad\qquad\qquad~
O_2 = (\bar u_\alpha b_\beta)_\vma(\bar q_\beta u_\alpha)_\vma,
\non \\
 && O_{3(5)}=(\bar sd)_\vma\sum_{q}(\bar qq)_{\vma(\vpa)}, \qquad
\qquad O_{4(6)}=(\bar s_\alpha d_\beta)_\vma\sum_{q}(\bar
q_\beta q_\alpha)_{ \vma(\vpa)},   \\
&& O_{7(9)}={3\over 2}(\bar sd)_\vma\sum_{q}e_{q}(\bar qq)_{\vpa(\vma)},
  \qquad~ O_{8(10)}={3\over 2}(\bar s_\alpha d_\beta)_\vma\sum_{q}e_{q}(\bar
q_\beta q_\alpha)_{\vpa(\vma)},  \non
\en
with $O_3$--$O_6$ being the QCD penguin operators,
$O_{7}$--$O_{10}$ the electroweak penguin operators and $(\bar
q_1q_2)_{_{V\pm A}}\equiv\bar q_1\gamma_\mu(1\pm \gamma_5)q_2$.
The sum in Eq. (\ref{Oi}) is over light flavors, $q=u,d,s$. It is
obvious that only the Wilson coefficients $z_i$ are relevant to
our purposes, as we are only interested in the CP-conserving part
of $K\to\pi\pi$ transitions.

In order to apply the factorization approximation to hadronic
matrix elements, we need to compute the $\mu$ dependence of matrix
elements arising from vertex and penguin-type radiative
corrections to four-quark operators and combine it with the Wilson
coefficients to form renormalization-scale and -scheme independent
effective Wilson coefficient functions (for details, see
\cite{CLY,CCTY}):
\be
z_1^{\rm eff} &=& z_1(\mu)+{\alpha_s\over
4\pi}\left(\gamma_V^{(0)T}\ln{\mu_f\over \mu}+\hat
r_V^T\right)_{1i}z_i(\mu), \non \\
z_2^{\rm eff}
&=& z_2(\mu)+{\alpha_s\over 4\pi}\left(\gamma_V^{(0)T}\ln{\mu_f\over
\mu}+\hat r_V^T\right)_{2i}z_i(\mu), \non \\
z_3^{\rm eff} &=& z_3(\mu)+{\alpha_s\over 4\pi}\left(\gamma_V^{(0)T}
\ln{\mu_f\over \mu}+\hat r_V^T\right)_{3i}z_i(\mu)-{\alpha_s
\over 24\pi}(C_t+C_p), \non \\
z_4^{\rm eff} &=& z_4(\mu)+{\alpha_s\over
4\pi}\left(\gamma_V^{(0)T}\ln{\mu_f\over \mu}+\hat
r_V^T\right)_{4i}z_i(\mu)+{\alpha_s\over 8\pi}(C_t+C_p), \non \\
z_5^{\rm eff} &=& z_5(\mu)+{\alpha_s\over
4\pi}\left(\gamma_V^{(0)T}\ln{\mu_f\over \mu}+\hat
r_V^T\right)_{5i}z_i(\mu)-{\alpha_s\over 24\pi}(C_t+C_p), \non \\
z_6^{\rm eff} &=& z_6(\mu)+{\alpha_s\over
4\pi}\left(\gamma_V^{(0)T}\ln{\mu_f\over \mu}+\hat
r_V^T\right)_{6i}z_i(\mu)+{\alpha_s\over 8\pi}(C_t+C_p), \non \\
z_7^{\rm eff} &=& z_7(\mu)+{\alpha_s\over
4\pi}\left(\gamma_V^{(0)T}\ln{\mu_f\over \mu}+\hat
r_V^T\right)_{7i}z_i(\mu)+{\alpha\over 8\pi}C_e, \non \\
z_8^{\rm eff} &=& z_8(\mu)+{\alpha_s\over
4\pi}\left(\gamma_V^{(0)T}\ln{\mu_f\over \mu}+\hat
r_V^T\right)_{8i}z_i(\mu), \non \\
z_9^{\rm eff} &=& z_9(\mu)+{\alpha_s\over 4\pi}\left(\gamma_V^{(0)T}
\ln{\mu_f\over
\mu}+\hat r_V^T\right)_{9i}z_i(\mu)+{\alpha\over 8\pi}C_e, \non \\
z_{10}^{\rm eff} &=& z_{10}(\mu)+{\alpha_s\over
4\pi}\left(\gamma_V^{(0)T}\ln{\mu_f\over \mu}+\hat
r_V^T\right)_{10i}z_i(\mu), \label{ceff1}
\en
where the superscript $T$ denotes a transpose of the matrix, the
anomalous dimension matrix $\gamma_V^{(0)}$ as well as the
constant matrix $\hat r_V$ arise from the vertex corrections to
the operators $O_1-O_{10}$, $C_t$, $C_p$ and $C_e$ from the QCD
penguin-type diagrams of the operators $O_{1,2}$, the QCD
penguin-type diagrams of the operators $O_3-O_6$, and the
electroweak penguin-type diagram of $O_{1,2}$, respectively:
\be
C_t &=& \tilde G(m_u)z_1, \non \\
C_p &=& [\tilde G(m_s)+\tilde G(m_d)]z_3+\sum_{i=u,d,s}\tilde
G(m_i)(z_4+z_6),  \non \\
C_e &=& {8\over 9}\tilde G(m_u)(z_1+3z_2),
\non \\
\tilde G(m_q) &=& {2\over 3}\kappa-G(m_q,k,\mu),
\label{ct}
\en
with $\kappa$ being a
parameter characterizing the $\gamma_5$ scheme dependence in
dimensional regularization, for example,
\be
\kappa=\cases{ 1 & NDR,  \cr 0 & HV,  \cr}
\en
in the naive dimensional regularization (NDR) and 't Hooft-Veltman
(HV) schemes for $\gamma_5$. The function $G(m,k,\mu)$ in Eq.
(\ref{ct}) is given by
\be
\label{G} G(m,k,\mu)=-4\int^1_0dx\,x(1-x)\ln\left(
{m^2-k^2x(1-x)\over \mu^2}\right),
\en
where $k^2$ is the momentum squared carried by the virtual gluon.

The matrix $\hat r$ in (\ref{ceff1}) gives momentum-independent
constant terms which depend on the treatment of $\gamma_5$ in
dimensional regularization. An early evaluation of $\hat r$ is
performed in the off-shell quark scheme \cite{Ali}. However, it
was pointed out by Buras and Silvestrini \cite{Buras98} that
$z_i^{\rm eff}$ thus constructed suffer from gauge and infrared
ambiguities since an off-shell external quark momentum, which is
usually chosen to regulate the infrared divergence occurred in the
radiative corrections to the local 4-quark operators, will
introduce a gauge dependence. It was shown recently in \cite{CLY}
that the above-mentioned problems on gauge dependence and infrared
singularity connected with the effective Wilson coefficients can
be resolved by perturbative QCD (PQCD) factorization theorem. In
this formalism, partons, {\it i.e.}, external quarks, are assumed
to be on shell, and both ultraviolet and infrared divergences in
radiative corrections are isolated using the dimensional
regularization. Because external quarks are on shell, gauge
invariance of the decay amplitude is maintained under radiative
corrections to all orders. This statement is confirmed by an
explicit one-loop calculation in \cite{CLY}. The obtained
ultraviolet poles are subtracted in a renormalization scheme,
while the infrared poles are absorbed into universal
nonperturbative bound-state wave functions. Explicitly, the
effective Wilson coefficient has the generic expression
\begin{eqnarray}
c^{\rm eff}=c(\mu)g_1(\mu)g_2(\mu_f)\;, \label{nef}
\end{eqnarray}
where $g_1(\mu)$ is an evolution factor from the scale $\mu$ to
$m_Q$, whose anomalous dimension is the same as that of $c(\mu)$,
and $g_2(\mu_f)$ describes the evolution from $m_Q$ to $\mu_f$
($\mu_f$ being a factorization scale arising from the dimensional
regularization of infrared divergences), whose anomalous dimension
differs from that of $c(\mu)$ because of the inclusion of the
dynamics associated with spectator quarks. For kaon decays under
consideration, there is no any heavy quark mass scale between
$m_c$ and $m_K$. Hence, the logarithmic term emerged in the vertex
corrections to 4-quark operators is of the form $\ln \mu_f/\mu$ as
shown in Eq. (\ref{ceff1}). We will set $\mu_f=1$ GeV in order to
have a reliable estimate of perturbative effects on effective
Wilson coefficients.

The scale dependence of vertex and penguin-type corrections shown
in Eq. (\ref{ceff1}) is governed by the terms
$\gamma_V^{(0)}\ln\mu$ and $\tilde G(\mu)$, while the
$\gamma_5$-scheme dependence is determined by the matrix $\hat
r_V$ as well as $\tilde G(\kappa)$. Formally, one can show that
the $\mu$ and $\gamma_5$-scheme dependence of the next-to-leading
order (NLO) Wilson coefficient, say $z_1(\mu)$, is compensated by
$\gamma_V^{(0)}\ln\mu$ and $\hat r_V$, respectively, to the order
of $\alpha_s$ \cite{CLY,CCTY}. This means that the NLO Wilson
coefficients $z_i(\mu)$ appearing in Eq. (\ref{ceff1}) together
with $\alpha_s/4\pi$ or $\alpha/4\pi$ should be replaced by the
lowest-order values $z_i^{\rm LO}(\mu)$. The numerical values of
$z_i^{\rm eff}$ are displayed in Table I. We see that except for
$z_{6,7}^{\rm eff}$, effective Wilson coefficients shown in the
last two columns of Table I are indeed renormalization scheme
independent, as it should be.

\vskip 0.4cm
\begin{table}[ht]
\caption{ $\Delta S=1$ Wilson coefficients at $\mu=1$ GeV for
$m_t=170$ GeV and $\Lambda_{\ov{\rm MS}}^{(4)}=325$ MeV, taken
from Table XVIII of [6]. Also shown are the effective Wilson
coefficients obtained from $z_i^{\rm NDR}(\mu)$ and $z_i^{\rm
HV}(\mu)$ via Eq. (2.4) with $\mu=1$ GeV, $\mu_f=1$ GeV and
$k^2=m_K^2/2$.}
\begin{center}
\begin{tabular}{ l r r r r r}
 & LO & NDR & HV & $z_i^{\rm eff}$(NDR) & $z_i^{\rm eff}$(HV)  \\ \hline
$z_1$ & 1.433 & 1.278 & 1.371 & 1.718 & 1.713 \\ $z_2$ & -0.748 &
-0.509 & -0.640 & -1.113 & -1.110 \\ $z_3$ & 0.004 & 0.013 & 0.007
& 0.034 & 0.033 \\ $z_4$ & -0.012 & -0.035 & -0.017 & -0.088 &
-0.087 \\ $z_5$ & 0.004 & 0.008 & 0.004 & 0.026 & 0.026 \\ $z_6$ &
-0.013 & -0.035 & -0.014 & -0.093 & -0.089 \\ $z_7/\alpha$ & 0.008
& 0.011 & -0.002 & 0.063 & 0.069 \\ $z_8/\alpha$ & 0.001 & 0.014 &
0.010 & 0.016 & 0.013 \\ $z_9/\alpha$ & 0.008 & 0.018 & 0.005 &
0.072 & 0.078 \\ $z_{10}/\alpha$ & -0.001 & -0.008 & -0.010 &
-0.011 & -0.012 \\
\end{tabular}
\end{center}
\end{table}

In the late 70's and early 80's, it had been suggested that
penguin operators may account for the $\Delta I=1/2$ rule observed
in kaon decays. With the advent of the effective Hamiltonian
approach, it is realized that the $\Delta I=1/2$ selection rule
cannot be dominated by the penguin mechanism. One popular argument
is that at the scale, say $\mu\gsim m_c$, the penguin Wilson
coefficients are negligible due to the (incomplete) GIM mechanism.
Since the penguin Wilson coefficients become important at $\mu=1$
GeV, for instance, one may wonder if the physical penguin
contributions to $K\to\pi\pi$ is independent of the choice of
$\mu$. The point is that although $z_3,\cdots,z_{10}$ vanish at,
say $\mu=2$ GeV, the effect of the penguin diagrams with the
internal $u$ quark induced by the current-current operator $O_1$
has to be taken into account when evaluating matrix elements.
Consequently, the total penguin contribution is scale independent,
and this is the merit of effective Wilson coefficients in which
the perturbative effect of the penguin diagram with the internal
$u$ quark is already included.

We add a remark before ending this section. A dynamical phase can
arise from the time-like penguin diagram involving internal $u$
loop quarks. However, since this penguin-induced phase is
incorporated into the isospin zero final-state interaction phase
shift $\delta_0$ to be introduced below in Eq. (\ref{A02}), its
contribution should not be double-counted in the effective Wilson
coefficients.

\section{Calculations}
In this section we first study the $K\to\pi\pi$ matrix elements
based on the vacuum insertion approximation, and then turn to the
$\Delta I=1/2$ and $\Delta I=3/2$ amplitudes.
\subsection{Matrix Elements}
It is convenient to make isospin decomposition of the matrix
elements $\la O_i\ra_{0,2}\equiv \la\pi\pi,I=0,2|O_i|K^0\ra$ which
are related to $K-\pi\pi$ transitions via
\be
\la O_i\ra_0 &=&
{1\over\sqrt{6}}\left(2\la\pi^+\pi^-|O_i|K^0\ra+\la\pi^0\pi^0
|O_i|K^0\ra\right),  \non \\ \la O_i\ra_2 &=&
{1\over\sqrt{3}}\left(\la\pi^+\pi^-|O_i|K^0\ra-\la\pi^0\pi^0
|O_i|K^0\ra\right)=\sqrt{2\over 3}\,\la\pi^+\pi^0|O_i|K^+\ra.
\en
Conventionally, the matrix elements $\la O_i\ra_{0,2}$ are
evaluated using the vacuum insertion approximation (i.e.
factorization hypothesis). Under this assumption, we have, for
example,
\be
\la\pi^+\pi^-|O_1|K^0\ra &=& \la\pi^+|(\bar ud)|0\ra\la\pi^-|(\bar
su)|K^0\ra+{1\over N_c}\la\pi^+\pi^-|(\bar uu)|0\ra\la 0|(\bar
sd)|K^0\ra  \non\\ &=& f_\pi(m_K^2-m_\pi^2)F_0^{K\pi}(m_\pi^2),
\en
where the form factor $F_0$ is defined in \cite{BSW85} and the
$W$-exchange contribution vanishes due to vector current
conservation. The $q^2$ dependence of the form factor $F_0$ is
usually assumed to be dominated by near poles in a monopole
manner:
\be
F_0^{K\pi}(q^2)={F_0^{K\pi}(0)\over 1-{q^2\over m^2_*}}\approx
F_0^{K\pi}(0)\left(1+{q^2\over m^2_*}\right), \label{m.e.1}
\en
where $m_*$ is the pole mass of the $0^+$ scalar meson with the
quantum number of $s\bar q$ ($q=u,d$). In chiral perturbation
theory (ChPT), we have $F_0^{K\pi}(0)=1$ due to vector current
conservation and (see \cite{Cheng89} for details)
\be
\label{Lambda} {1\over m_*^2}={8L_5\over f_\pi^2}=\left({f_K\over
f_\pi}-1\right){1\over (m^2_K-m^2_\pi)}\approx
{1\over\Lambda^2_\chi},
\en
where $L_5$ is one of the coupling constants in the ${\cal
O}(p^4)$ chiral Lagrangian for strong interactions, and
$\Lambda_\chi\approx 2\pi f_\pi$ (our $f_\pi=132$ MeV) is the
chiral-symmetry breaking scale \cite{Cheng86}. Therefore,
\be
\la\pi^+\pi^-|O_1|K^0\ra=\,f_\pi(m^2_K-m_\pi^2)\left(1+{m^2_\pi\over
\Lambda^2_\chi}\right).
\en
In ChPT, the term proportional to $m_\pi^2/\Lambda^2_\chi$ is
counted as a contribution of ${\cal O}(p^4)$.

Contrary to charmless $B$ decays, the penguin operators $O_{5,6}$,
do not contribute directly to $K^0\to\pi\pi$ because of the wave
function $\pi^0={1\over\sqrt{2}}(\bar uu-\bar dd)$ and the
SU(3)-singlet nature of the $V+A$ current. Nevertheless, the Fierz
transformation of $O_{5,6}$ via $(V-A)(V+A)\to -2(S+P)(S-P)$ does
make contributions. For example,
\be
\la\pi^+\pi^-|O_6|K^0\ra &=& {2\over 3}\la\pi^+|\bar
u\gamma_5d|0\ra\la\pi^-|\bar s u|K^0\ra-{2\over
3}\la\pi^+\pi^-|\bar dd|0\ra\la 0|\bar s\gamma_5 d|K^0\ra \non \\
&+& {2\over 3}(\la\bar dd\ra+\la\bar ss\ra)\la\pi^+\pi^-|\bar
s\gamma_5d|K^0\ra.  \label{o6me}
\en
The matrix elements of scalar and pseudoscalar densities can be
evaluated using equations of motion or the chiral representation
of quark densities. The former method gives
\be
\la\pi^-(q)|\bar su|K^0(k)\ra=\,v\left[ 1+{(k-q)^2\over
\Lambda^2_\chi}\right], && \qquad\quad \la\pi^+(q)|\bar
u\gamma_5d|0\ra=if_\pi v, \non \\ \la\pi^+(q_+)\pi^-(q_-)|\bar
dd|0\ra=\,v\left[ 1+{(q_++q_-)^2\over \Lambda^2_\chi}\right],
 && \qquad\quad \la 0|\bar s\gamma_5d|K^0(k)\ra=if_K v,
\en
where uses of Eqs. (\ref{m.e.1}) and (\ref{Lambda}) have been
made, and
\be
v={m^2_{\pi^\pm}\over m_u+m_d}={m^2_{K^0}\over
m_d+m_s}={m^2_K-m^2_\pi\over m_s-m_u}
\en
characterizes the quark-order parameter $\la \bar qq\ra$ which
breaks chiral symmetry spontaneously. The second term on the
r.h.s. of Eq. (\ref{o6me}) is the so-called spacelike penguin
contribution. Unlike the case of hadronic charmless $B$ decays,
the spacelike penguin diagram in $K\to\pi\pi$ is calculable.  The
last term in Eq. (\ref{o6me}), which is a tadpole contribution
arising from the vacuum expectational values of quark bilinears,
does not contribute to the physical $K\to\pi\pi$ amplitude
\cite{Cheng89}. Hence, we obtain
\be
\la\pi^+(q_+)\pi^-(q_-)|O_6|K^0(k)\ra=-i{4\over 3}f_\pi
v^2\,{k^2-q_+^2\over \Lambda^2_\chi}+{\cal O}\left({1\over
\Lambda^4_\chi}\right),
\en
where we have applied Eq. (\ref{Lambda}).

The matrix elements obtained under the vacuum insertion
approximation are summarized below:
\be
\la O_1\ra_0 &=& {1\over 3}X\left(2-{1\over N_c}\right), \qquad
\qquad \quad \la O_1\ra_2 = {\sqrt{2}\over 3}X\left(1+{1\over
N_c}\right), \non \\ \la O_2\ra_0 &=& {1\over 3}X\left(-1+{2\over
N_c}\right), \qquad \qquad~ \la O_2\ra_2 = {\sqrt{2}\over
3}X\left(1+{1\over N_c}\right), \non \\ \la O_3\ra_0 &=& {1\over
N_c}X, \qquad\qquad \qquad\qquad\quad \la O_4\ra_0=X, \non \\ \la
O_5\ra_0 &=& -{4\over N_c}\,\sqrt{3\over 2}\,v^2(f_K-f_\pi),
\qquad~ \la O_6\ra_0 = -4\,\sqrt{3\over 2}\,v^2(f_K-f_\pi), \non\\
\la O_7\ra_0 &=& {\sqrt{6}\over N_c}f_K v^2+{1\over 2}X, \qquad
\qquad \quad \la O_7\ra_2 = {\sqrt{3}\over N_c}f_\pi v^2-{1\over
\sqrt{2}}X, \non\\ \la O_8\ra_0 &=& \sqrt{6}f_K v^2+{1\over
2N_c}X, \qquad \qquad \la O_8\ra_2 = \sqrt{3}f_\pi v^2-{1\over
N_c\sqrt{2}}X, \non\\ \la O_9\ra_0 &=& -{1\over 2}X\left(1-{1\over
N_c}\right), \qquad\qquad~ \la O_9\ra_2 = -{1\over
\sqrt{2}}X\left(1+{1\over N_c}\right), \non \\ \la O_{10}\ra_0 &=&
{1\over 2}X\left(1-{1\over N_c}\right), \qquad \qquad\quad \la
O_{10}\ra_2 = {1\over \sqrt{2}}X\left(1+{1\over N_c}\right),
\label{O02me}
\en
where $X=\sqrt{3/2}\,f_\pi(m_K^2-m_\pi^2)$ and $1/\Lambda^2_\chi$
corrections to $(V-A)(V\pm A)$ matrix elements as well as
$1/\Lambda^4_\chi$ corrections to $(S+P)(S-P)$ matrix elements
have been neglected.

In terms of the isospin matrix elements, the corresponding isospin
decay amplitudes are given by
\be
A_Ie^{i\delta_I} &=& {G_F\over\sqrt{2}}V_{ud}V_{us}^*\sum_i
c_i(\mu)\la O_i(\mu)\ra_I,  \non \\
 &=& {G_F\over\sqrt{2}}V_{ud}V_{us}^*\sum_i
c_i^{\rm eff}\la O_i\ra_I, \label{A02}
\en
and hence
\be
{\rm Re}\,A_{0,2} =
{G_F\over\sqrt{2}}\,{V_{ud}V_{us}^*\over\cos\delta_{0,2}}\sum_i
z_i^{\rm eff}\la O_i\ra_{0,2}, \label{ReA}
\en
where $\delta_0$ and $\delta_2$ are S-wave $\pi\pi$ scattering
isospin phase shifts. In the present paper, we will use the
analysis of \cite{Chell} for phase shifts:
\be
\delta_0=(34.2\pm 2.2)^\circ, \qquad\quad \delta_2=-(6.9\pm
0.2)^\circ.
\en
Experimentally, the isospin $K\to\pi\pi$ amplitudes are given by
\cite{Cheng89}
\be
{\rm Re}\,A_0=3.323\times 10^{-7}\,{\rm GeV}, \qquad {\rm Re}\,
A_2=1.497\times 10^{-8}\,{\rm GeV}. \label{expt}
\en

\subsection{The $\Delta I=3/2$ amplitude}
It is straightforward to show from Eqs. (\ref{O02me}) and
(\ref{ReA}) that the isospin 2 amplitude of $K\to\pi\pi$ has the
form
\be
A_2^{(0)} =
{G_F\over\sqrt{2}}\,{V_{ud}V^*_{us}\over\cos\delta_2}\Bigg\{
\Big[a_1+a_2+{3\over 2}(-a_7+a_9+a_{10})\Big]\,{\sqrt{2}\over 3}
X+\sqrt{3}\,f_\pi v^2a_8\Bigg\}, \label{A20}
\en
where
\be
a_{2i}= {z}_{2i}^{\rm eff}+{1\over N_c}{z}_{2i-1}^{\rm eff},
\qquad\quad a_{2i-1}= {z}_{2i-1}^{\rm eff}+{1\over N_c}{z}^{\rm
eff}_{2i}  \label{ai}
\en
for $i=1,\cdots,5$, and the superscript (0) indicates that this
amplitude is induced by pure $\Delta I=3/2$ weak interactions.
Since the electroweak penguin coefficients are very small compared
to $z_1^{\rm eff}$ and $z_2^{\rm eff}$, it is clear that the
$\Delta I=3/2$ decay amplitude is entirely governed by
current-current 4-quark operators. It is also known that
$K^+\to\pi^+\pi^0$ (or $\Delta I=3/2$ $K^0\to\pi\pi$ decays) can
be generated from the $\Delta I=1/2$ decays
$K^+\to\pi^+\eta\,(\eta')$ followed by the isospin breaking mixing
$\pi^0-\eta-\eta'$ \cite{Holstein,Cheng88}. As a result, the total
$\Delta I=3/2$ amplitude reads
\be
A_2=\,{A_2^{(0)}\over 1-\Omega_{\rm IB}}, \label{A2total}
\en
where the expression of $\Omega_{\rm IB}\equiv A_2^{\rm IB}/A_2$
can be found in Appendix A . Employing the quark mass ratios
$m_d/m_u=0.553\pm 0.043$ and $m_s/m_d=18.9\pm 0.8$ obtained in a
recent detailed analysis based on ChPT \cite{Leutwyler}, we find
from Eq. (\ref{Omega}) that
\be
\Omega_{\rm IB}=0.25\pm 0.02\,. \label{vomega}
\en

Eqs. (\ref{A20})-(\ref{vomega}) lead to
\be
A_2=4.133\,(z_1^{\rm eff}+z_2^{\rm eff})\left(1+{1\over
N_c}\right)\times 10^{-8}{\rm GeV}. \label{A2theory}
\en
Using the effective Wilson coefficients $z_i^{\rm eff}$ given in
Table I, it is easily seen that the predicted $A_2$ is too large
by a factor of 2.2 compared to experiment (\ref{expt}). This means
that nonfactorized contributions that have been neglected thus far
should be taken into account. For $K\to\pi\pi$ decays,
nonfactorizable effects in hadronic matrix elements can be
absorbed into the parameters $a_i^{\rm eff}$
\cite{Cheng94,Kamal94,Soares}:
\be
a_{2i}^{\rm eff}= {z}_{2i}^{\rm eff}+\left({1\over
N_c}+\chi_{2i}\right){z}_{2i-1}^{\rm eff}, \qquad\quad
a_{2i-1}^{\rm eff}= {z}_{2i-1}^{\rm eff}+\left({1\over
N_c}+\chi_{2i-1}\right){z}^{\rm eff}_{2i}, \label{aeff}
\en
where the nonfactorized terms $\chi_{1,2}$ relevant to
$K^+\to\pi^+\pi^0$ decay are given by
\cite{Soares,Kamal96,Neubert}
\be
\chi_1 =\vp_8^{(K^+\pi^0,\pi^+)}+{a_1\over z_2^{\rm eff}
}\vp_1^{(K^+\pi^0,\pi^+)},\qquad \chi_2
=\vp_8^{(K^+\pi^+,\pi^0)}+{a_2\over z_1^{\rm eff}
}\vp_1^{(K^+\pi^+,\pi^0)}, \label{chi}
\en
with $a_{1,2}=z_{1,2}^{\rm eff}+z_{2,1}^{\rm eff}/N_c$, and
\be
\vp_1^{(K^+\pi^0,\pi^+)} &=& {\la \pi^+\pi^0|(\bar ud)_\vma(\bar
su)_\vma|K^+ \ra_{ nf}\over \la \pi^+\pi^0|(\bar ud)_\vma(\bar
su)_\vma|K^+\ra_f}= {\la \pi^+\pi^0|(\bar ud)_\vma(\bar
su)_\vma|K^+ \ra\over \la \pi^+|(\bar ud)_\vma|0\ra\la \pi^+|(\bar
su)_\vma|K^+\ra}-1, \non\\ \vp_8^{(K^+\pi^0,\pi^+)} &=& {1\over
2}\,{\la \pi^+\pi^0|(\bar u\lambda^a d)_\vma (\bar s\lambda^a
u)_\vma|K^+ \ra\over \la \pi^+|(\bar ud)_\vma|0\ra\la \pi^0|(\bar
su)_\vma|K^+\ra}, \label{vp}
\en
being nonfactorizable terms originated from color-singlet and
color-octet currents, respectively, and $(\bar q_1\lambda^a
q_2)_\vma\equiv \bar q_1\lambda^a \gamma_\mu (1-\gamma_5)q_2$. The
subscripts `f' and `nf' in Eq.~(\ref{vp}) stand for factorizable
and nonfactorizable contributions, respectively, and the
superscript $(K^+\pi^0,\pi^+)$ in Eq.~(\ref{chi}) means that the
$\pi^+$ is factored out in the factorizable amplitude of
$K^+\to\pi^+\pi^0$ and likewise for the superscript
$(K^+\pi^+,\pi^0)$. In the large-$N_c$ limit, $\vp_1={\cal
O}(1/N_c^2)$ and $\vp_8={\cal O}(1/N_c)$ \cite{Neubert}.
Therefore, the nonfactorizable term $\chi$ in the $N_c\to \infty$
limit is dominated by color octet-octet operators.

Assuming $\chi_1=\chi_2$ in Eq. (\ref{aeff}) for $a_1^{\rm eff}$
and $a_2^{\rm eff}$ and fitting Eq. (\ref{A2theory}), in which
$1/N_c$ is replaced by $1/N_c+\chi$, to the experimental value
(\ref{expt}), we obtain\footnote{In the so-called large-$N_c$
approach, one has $\chi=-1/3$ to the leading $1/N_c$ expansion.}
\be
\chi(K\to\pi\pi)=-0.73\,,
\en
and hence
\be
a_1^{\rm eff}=2.16\,, \qquad\qquad a_2^{\rm eff}=-1.80\,.
\label{aeff12}
\en
For comparison, the nonfactorized effects in hadronic two-body
decays of charmed and bottom mesons are given by \cite{CK99}
\be
\chi_2(D\to \overline{K}\pi)\sim -0.33\,, \qquad \chi_2(B\to
D\pi)\sim (0.12-0.21)\,.
\en
The fact that
\be
|\chi(K\to\pi\pi)|\gg|\chi_2(D\to \ov K\pi)|\gg|\chi_2(B\to D\pi
)|
\en
is consistent with the intuitive picture that soft gluon effects
become stronger when final-state particles move slower, allowing
more time for significant final-state interactions after
hadronization \cite{Cheng94}.

Note that in $B$ or $D$ decays, the parameters $a_{1,2}^{\rm eff}$
and hence $\chi_{1,2}$  in principle can be determined separately
from experiments under some plausible assumptions. For example,
$\chi_1(D\to\ov K\pi)$ and $\chi_2(D\to\ov K\pi)$ can be extracted
from the isospin analysis of $D^0\to K^-\pi^+,\,\ov K^0\pi^0$ and
$D^+\to\ov K^0\pi^+$ data provided that the $W$-exchange is
negligible.\footnote{Since in general $|z_1/z_2|\gg 1$, the
determination of $\chi_2$ is easier and more reliable than
$\chi_1$.} By contrast, $\chi_1(K\to\pi\pi)$ and
$\chi_2(K\to\pi\pi)$ cannot be determined from the data without
invoking a further assumption because neutral $K^0\to\pi\pi$
decays receive additional penguin contributions. That is why we
make the universality assumption $\chi_1=\chi_2$ to extract
$a_{1,2}(K\to\pi\pi)$ from the measurement of $K^+\to\pi^+\pi^0$.

In the literature, the effective parameters $a_i^{\rm eff}$ are
sometimes expressed in terms of the the scheme- and
scale-dependent Wilson coefficients $z_i(\mu)$, for example,
\be
a_{1}^{\rm eff}= {z}_{1}(\mu)+\left({1\over
N_c}+\tilde\chi_{1}(\mu)\right){z}_{2}(\mu), \qquad\quad
a_{2}^{\rm eff}= {z}_{2}(\mu)+\left({1\over
N_c}+\tilde\chi_{2}(\mu)\right){z}_{1}(\mu),
  \label{a12eff}
\en
where we have put a tilde on $\chi_{1,2}$ to distinguish them from
$\chi_{1,2}$ defined in Eq. (\ref{aeff}). Then it is clear that
$\tilde\chi_{1,2}$  must be $\gamma_5$-scheme and scale dependent
in order to ensure the scheme and scale independence of
$a_{1,2}^{\rm eff}$. Comparing Eqs. (\ref{a12eff}) and
(\ref{aeff}), we see that $\tilde\chi_{1,2}$ receive contributions
from vertex radiative corrections. It should be stressed that the
assumption $\tilde\chi_1=\tilde
\chi_2$ cannot lead to
$\gamma_5$-scheme independent $a_{1,2}^{\rm eff}$. To see this, we
assume (\ref{aeff12}) to be the true values for $a_{1,2}^{\rm
eff}$ and apply the Wilson coefficients evaluated at $\mu=1 $ GeV
in NDR and HV schemes given in Table I.  We find
\be
\tilde\chi_1^{\rm NDR}(\mu)=-2.07\,, &&\qquad\qquad
\tilde\chi_2^{\rm NDR}(\mu)=-1.34\,, \non\\ \tilde\chi_1^{\rm
HV}(\mu)=-1.57\,, &&\qquad\qquad \tilde\chi_2^{\rm
HV}(\mu)=-1.18\,, \label{chi12}
\en
at $\mu=1$ GeV by fitting (\ref{a12eff}) to (\ref{aeff12}). This
implies that phenomenologically it is not possible to determine
$a_{1,2}^{\rm eff}$ from the data of $K\to \pi\pi$ if we start
with the scheme- and scale-dependent Wilson coefficients
$z_i(\mu)$ without taking into account vertex corrections to
$\tilde\chi_{1,2}$.

\subsection{The $\Delta I=1/2$ amplitude}
From Eqs. (\ref{O02me}) and (\ref{ReA}) we obtain the $\Delta
I=1/2$ amplitude:
\be
A_0 &=&
{G_F\over\sqrt{2}}\,{V_{ud}V^*_{us}\over\cos\delta_0}\Bigg\{
\Big[\,{2\over 3}a_1-{1\over 3}a_2+a_4+{1\over
2}(a_7-a_9+a_{10})\Big]X  \non \\ &-&
2\sqrt{6}\,v^2(f_K-f_\pi)a_6+\sqrt{6}\,v^2f_K a_8\Bigg\},
\label{A0}
\en
where we have neglected the contribution arising from
$\pi^0-\eta-\eta'$ mixing. For simplicity, we have also dropped
the superscript `eff' of the parameters $a_i$. To incorporate
nonfactorized effects, we shall make the universality assumption:
\be
&& \chi_{LL}\equiv\chi_1=\chi_2=\chi_3=\chi_4=\chi_9=\chi_{10},
\non
\\ && \chi_{LR}\equiv\chi_5=\chi_6=\chi_7=\chi_8.
\en
The nonfactorized effects in the matrix elements of $(V-A)(V+A)$
operators are {\it a priori} different from that of $(V-A)(V-A)$
operators. Indeed, we have learned from hadronic charmless $B$
decays that $\chi_{LR}\neq \chi_{LL}$ \cite{CCTY}. However, in the
absence of information for the nonfactorized contributions to
$K\to \pi\pi$ penguin operators, we shall assume
$\chi_{LR}\approx\chi_{LL}=-0.73$ for simplicity. Moreover, we
found in actual calculations, $A_0$ is insensitive to the value of
$\chi_{LR}$. From Eq. (\ref{A0}) and Table I, it is easily seen
that the nonfactorized term $\chi_{LL}=-0.73$, which is needed to
suppress $A_2$ to the observed value, will enhance the tree
contribution to $A_0$ by a factor of 1.9; that is, the tree
contribution to $A_0/A_2$ ratio is increased by a factor of 3 !

Treating the strange quark mass $m_s$ and hence the parameter $v$
as a free parameter, we plot in Fig. 1 the ratio $A_0/A_2$ as a
function of $m_s$ at the renormalization scale $\mu=1$ GeV.
Specifically, we obtain
\be
{A_0\over A_2}=\cases{ 17.1 & at~$m_s\,(1\,{\rm GeV})=127$ MeV,
\cr 15.3 & at~$m_s\,(1\,{\rm GeV})=150$ MeV.}
\en
It is clear that $m_s$ is favored to be smaller. Presently there
is no consensus regarding the values of light quark masses. It is
interesting to note that several recent lattice calculations give
a lighter strange quark mass: Results using the
Sheikholeslami-Wolhlert fermion yield $m_s=(95\pm 16)$ MeV
\cite{Gough}, a computation based on domain wall fermions obtains
$m_s=(95\pm 26)$ MeV \cite{Soni}, a quenched QCD calculation
together with the quark mass ratios from ChPT gives $m_s=(97\pm
7)$ MeV \cite{Garden}, and a new unquenched lattice result
indicates a still lower number $m_s=(84\pm 7)$ MeV \cite{Ryan},
all in the $\overline{\rm MS}$ scheme at a scale of 2 GeV. The
strange quark mass 95 MeV at $\mu=1$ GeV corresponds to $m_s=127$
MeV at $\mu=1$ GeV.

\begin{figure}[tb]
\psfig{figure=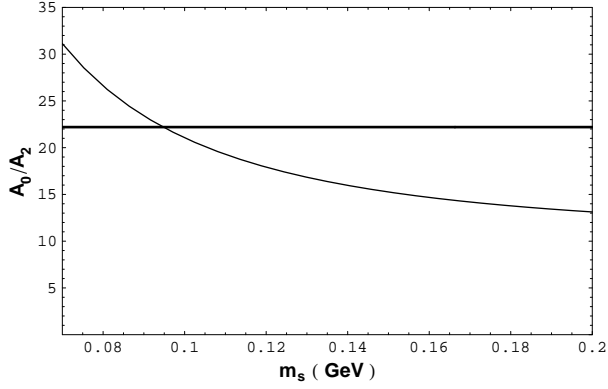,height=2.0in} \vspace{0.4cm}
    \caption{{\small The ratio of $A_0/A_2$ versus $m_s$ (in units of
    GeV) at the renormalization scale $\mu=1$ GeV. The solid thick line
    is the experimental value for $A_0/A_2$.}}
\end{figure}

It is instructive to see how the prediction of the $\Delta I=1/2$
rule progresses at various steps. In the absence of QCD
corrections, we have $a_2={1\over 3}a_1$ and
$a_3=a_4\cdots=a_{10}=0$ under the vacuum insertion approximation.
It follows from Eqs. (\ref{A20}) and (\ref{A0}) that
\cite{Cheng89}
\be
{A_0\over A_2}=\,{5\over 4\sqrt{2}}=0.9 \quad{\rm (in~
absence~of~QCD~corrections)}. \label{naive}
\en
With the inclusion of lowest-order short-distance QCD corrections
to the Wilson coefficients $z_1$ and $z_2$ evaluated at $\mu=1$
GeV, $A_0/A_2$ is enhanced from the value of 0.9 to 2.0, and it
becomes 2.4 for $m_s(1\,{\rm GeV})=127$ MeV when QCD and
electroweak penguin effects are included. This ratio is suppressed
to 1.8 with the inclusion of the isospin-breaking effect, but it
is increased again to the value of 2.1 in the presence of
final-state interactions with $\delta_0=34.2^\circ$ and
$\delta_2=-6.9^\circ$. Replacing $c_i^{\rm LO}(\mu)$ by the
effective Wilson coefficients $c_i^{\rm eff}$, or equivalently
replacing the LO Wilson coefficients by the NLO ones and including
vertex and penguin-type corrections to four-quark operators, we
find $A_0/A_2=4.8$\,. Finally, the inclusion of nonfactorized
effects on hadronic matrix elements will enhance $A_0/A_2$ to the
value of 17.1\,. In short, the enhancement of the ratio $A_0/A_2$
is due to the cumulative effects of the short-distance Wilson
coefficients, penguin operators, final-state interactions,
nonfactorized effects due to soft-gluon exchange, and radiative
corrections to the matrix elements of four-quark operators. Among
them, the last two effects, which are usually not addressed in
previous studies (in particular, the last one), play an essential
role for explaining the bulk of the $\Delta I=1/2$ rule. In
present calculations, penguin operators account for 35\% of the
$\Delta I=1/2$ rule for $m_s(1\,{\rm GeV})=127$ MeV.

Note that thus far we have neglected the $W$-exchange effect,
vanishing in he vacuum insertion approximation. Since the
$W$-exchange amplitude in charmed meson decay is comparable to the
internal $W$-emission one \cite{CC}, it is conceivable that in
kaon physics the long-distance contribution to $W$-exchange is as
important as the external $W$-emission amplitude. Therefore, the
$W$-exchange mechanism could provide an additional important
enhancement of the $A_0/A_2$ ratio.

\section{Comparison with Chiral Approach}
Since the scale dependence of hadronic matrix elements is lost in
the factorization approach, it has been advocated that a physical
cutoff $\Lambda_c$, which is introduced to regularize the
quadratic (and logarithmic) divergence of the long-distance chiral
loop corrections to $K\to\pi\pi$ amplitudes, can be identified
with the renormalization scale $\mu$ of the Wilson coefficients
\cite{BBG}. A most recent calculation along this line which
includes ${\cal O}(p^4)$ tree contributions \cite{Hambye}
indicates that while the isospin amplitude $A_0$ is largely
enhanced, the amplitude $A_2$ is highly unstable relative to the
cutoff scale $\Lambda_c$ and it even changes sign at
$\Lambda_c\gsim 650$ MeV \cite{Hambye,Bijnens}. The large
uncertainty for $A_2$ arises from the fact that the two
numerically leading terms, the tree level and the one-loop
quadratically divergent term, have approximately the same size but
opposite sign.

Since the scale dependence of Wilson coefficients is of the
logarithmic type, it seems quite unnatural to match the quadratic
cutoff with the $\mu$ dependence of $c(\mu)$. Therefore, it is
necessary to use the dimensional regularization to regularize the
chiral loop divergences and apply the same renormalization scheme,
say the $\ov{\rm MS}$ scheme, in order to consistently match the
scale dependence of Wilson coefficients evaluated using the same
regularization scheme. In this case, the inclusion of chiral loops
will make a large enhancement for $A_0$ and a small enhancement
for $A_2$ \cite{Kambor}. However, as stressed in passing, the
naive prediction of $A_2$ in the absence of nonfactorizable
effects is too large (by a factor 1.6 in our case) compared to
experiment. Therefore, chiral-loop corrections to $A_2$ will make
the discrepancy between theory and experiment even worse.
Evidently, this indicates that {\it not all the long-distance
nonfactorized contributions to hadronic matrix elements are fully
accounted for by chiral loops.} Several authors
\cite{Antonelli,Bijnens} have considered different models, for
instance the chiral quark model or the Nambu-Jona-Lasinio model,
to incorporate nonfactorized contributions arising from soft
gluonic corrections. For example, the nonfactorized gluonic
corrections computed in the chiral quark model amount to replacing
$1/N_c$ in the matrix elements $\la O_{1-3}\ra_{0,2}$ [see Eq.
(\ref{O02me})] by \cite{Antonelli}
\be
{1\over N_c}\to {1\over N_c}(1-\delta_{\la GG\ra})\equiv{1\over
N_c}\left( 1-{N_c\over 2}\,{4\pi^2\la\alpha_s GG/\pi\ra\over
\Lambda_\chi^4}\right),
\en
parametrized in terms of the gluon  condensate $\la\alpha_s
GG/\pi\ra$. It is clear that the correction $-\delta_{\la
GG\ra}/3$ plays the same role as the nonfactorized terms $\chi_i$
defined in Eq. (\ref{aeff}). This nonfactorized effect is
important since it can suppress the $A_2$ amplitude. Using
$\la\alpha_s GG/\pi\ra=(334\pm 4\,{\rm MeV})^4$ \cite{Bertolini},
one obtains $\delta_{\la GG\ra}=1.51$. It is clear that the soft
gluon correction, corresponding to $\chi_1=\chi_2=-0.50$, is large
enough to revert the sign of the $1/N_c$ term and thus suppress
the $A_2$ amplitude.

In the present analysis, the nonfactorized contribution to the
matrix elements of $(V-A)(V-A)$ operators characterized by the
nonfactorized terms $\chi_{1,2}=-0.73$ is comparable to that
obtained in the chiral quark model. Therefore, in a rough sense,
vertex and penguin-type radiative corrections to $K\to\pi\pi$
matrix elements in the effective Hamiltonian approach corresponds
to chiral-loop contributions in the aforementioned chiral
approach. However, the penguin contribution to the $A_0$ amplitude
in the latter approach is usually smaller than that in the former.
For example, the $O_6$ operator contribution to $A_0$ is about
20\% in \cite{Antonelli}, and it is even smaller in other
chiral-loop calculations.

\section{Conclusion}
We have studied $K\to\pi\pi$ decays within the framework of
generalized factorization in which the effective Wilson
coefficients are renormalization-scale and -scheme independent
while factorization is applied to the tree-level hadronic matrix
elements. Nonfactorizable contributions to the hadronic matrix
elements of $(V-A)(V-A)$ four-quark operators are extracted from
the measured $K^+\to\pi^+\pi^0$ decay to be $\chi_{1,2}=-0.73$
which explains the suppression of the $\Delta I=3/2$ $K\to\pi\pi$
amplitude $A_2$ and the enhancement of the $\Delta I=1/2$ $A_0$
amplitude. The $\Delta I=1/2$ rule arises from the cumulative
effects of the short-distance Wilson coefficients, penguin
operators, final-state interactions, nonfactorized effects due to
soft-gluon exchange, and radiative corrections to the matrix
elements of four-quark operators. In particular, the last two
effects are the main ingredients for the large enhancement of
$A_0$ with respect to $A_2$. The $A_0/A_2$ ratio is predicted to
lie in the range 15-17 for $m_s(1\,{\rm GeV})=(127-150)$ MeV.
Comparison of the present analysis of the $\Delta I=1/2$ rule with
the chiral-loop approach is given.

\vskip 1.0cm \acknowledgements  I wish to thank S. Bertolini for
useful comments. This work is supported in part by the National
Science Council of the Republic of China under Grants No.
NSC88-2112-M006-013.

\newpage
\centerline{\bf APPENDIX}
\renewcommand{\thesection}{\Alph{section}}
\renewcommand{\theequation}{\thesection\arabic{equation}}
\setcounter{equation}{0} \setcounter{section}{0} \vskip 0.5 cm
\vskip 0.3cm
\section{Isospin breaking effects on $\Delta I=3/2$ $K\to\pi\pi$
amplitude} In this Appendix we give an updated estimate of isospin
breaking contribution to $K^+\to\pi^+\pi^0$ due to the
$\pi-\eta-\eta'$ mixing. Writing
\be
A_2=A_2^{(0)}+A_2^{\rm IB},
\en
we have \cite{Cheng88}
\be
\Omega_{\rm IB}\equiv{A_2^{\rm IB}\over A_2}&=&{1\over
3\sqrt{2}}\,{A_0\over A_2}\,{m_d-m_u\over
m_s}\Bigg[(\cos\theta-\sqrt{2}\sin\theta)(\cos\theta-\sqrt{2}\,{\rho\over
1+\delta}\sin\theta)   \non \\ &+&
(\sin\theta+\sqrt{2}\cos\theta)(\sin\theta+\sqrt{2}\,{\rho\over
1+\delta}\cos\theta)\,{m_\eta^2-m^2_\pi\over
m^2_{\eta'}-m^2_\pi}\Bigg],
\en
where $\theta$ is the $\eta-\eta'$ mixing angle and the parameters
$\rho$ and $\delta$, defined by
\be
\la \eta_8|H_W|K^0\ra &=& \sqrt{1\over 3}(1+\delta)\la
\pi^0|H_W|K^0\ra,  \non \\ \la \eta_0|H_W|K^0\ra &=&
-2\sqrt{2\over 3}\,\rho\la \pi^0|H_W|K^0\ra,
\en
measure the breakdown of nonet symmetry in $K^0-\eta_0$ transition
and of SU(3)-flavor symmetry in $K^0-\eta_8$, respectively. We can
use the radiative decays $K_L\to\gamma\gamma$ and
$\pi^0\to\gamma\gamma$ to constrain $\rho$ and $\delta$:
\be
{A(K_L\to\gamma\gamma)\over
A(\pi^0\to\gamma\gamma)}=-4\,{m_K^2\over m_K^2-m^2_\pi}\,{g_8\over
f_\pi^2}\,\zeta,
\en
where
\be
\zeta &=& 1+{m^2_K-m^2_\pi\over m^2_K-m^2_\eta}\left(\sqrt{1\over
3}(1+\delta)\cos\theta+2\sqrt{2\over
3}\rho\sin\theta\right)\left(\sqrt{1\over 3}{f_\pi\over
f_8}\cos\theta-2\sqrt{2\over 3}{f_\pi\over f_0}\sin\theta\right)
\non \\ && +{m^2_K-m^2_\pi\over
m^2_K-m^2_{\eta'}}\left(\sqrt{1\over
3}(1+\delta)\sin\theta-2\sqrt{2\over
3}\rho\cos\theta\right)\left(\sqrt{1\over 3}{f_\pi\over
f_8}\sin\theta+2\sqrt{2\over 3}{f_\pi\over f_0}\cos\theta\right),
\en
and $g_8=0.26\times 10^{-5}m_K^2$ is the coupling constant in the
$\Delta S=1$ effective chiral Lagrangian.  From the data of
$K_L\to\gamma\gamma$ and $\pi^0\to\gamma\gamma$, we find
$|\zeta|=0.87$. Using
\be
\theta=-15.4^\circ, \qquad f_8/f_\pi=1.26,  \qquad f_0/f_\pi=1.17
\en
determined phenomenologically \cite{Kroll} and $\delta=0.17$
\cite{Donoghue}, we obtain $\rho=0.96$. Numerically, we find that
$\Omega_{\rm IB}$ is almost insensitive to the values of
$\delta,\,\rho$ and $\theta$ as long as they are constrained by
$\zeta$. Hence, to a very good approximation, we obtain
\be
\Omega_{\rm IB}=(10.45\pm 0.05)\,{m_d-m_u\over m_s}, \label{Omega}
\en
where the experimental value of $A_0/A_2=22.2$ has been used.

\section{Anomalous dimensional and constant matrices}
For reader's convenience, we list here the anomalous dimensional
matrix $\gamma_V^{(0)}$ and the constant matrix $\hat r_V$
appearing in Eq. (\ref{ceff1}):
\be
\label{gamma} \gamma_V^{(0)}=\left(\matrix{ -2 & 6 & 0 & 0 & 0 & 0
& 0 & 0 & 0 & 0 \cr 6 & -2 & 0 & 0 & 0 & 0 & 0 & 0 & 0 & 0 \cr 0 &
0 & -2 & 6 & 0 & 0 & 0 & 0 & 0 & 0 \cr  0 & 0 & 6 & -2  & 0 & 0 &
0 & 0 & 0 & 0 \cr 0 & 0 & 0 & 0 & 2 & -6 & 0 & 0 & 0 & 0 \cr 0 & 0
& 0 & 0 & 0 & -16 & 0 & 0 & 0 & 0 \cr 0 & 0 & 0 & 0 & 0 & 0 & 2 &
-6 & 0 & 0 \cr  0 & 0 & 0 & 0 & 0 & 0 & 0 & -16 & 0 & 0 \cr 0 & 0
& 0 & 0 & 0 & 0 & 0 & 0 & -2 & 6 \cr  0 & 0 & 0 & 0 & 0 & 0 & 0 &
0 & 6 & -2 \cr}\right),
\en
and \cite{CCTY}
\be
\label{rndr} \hat r_V^{\rm NDR}=\left(\matrix{ 3 & -9 & 0 & 0 & 0
& 0 & 0 & 0 & 0 & 0 \cr -9 & 3 & 0 & 0 & 0 & 0 & 0 & 0 & 0 & 0 \cr
0 & 0 & 3 & -9 & 0 & 0 & 0 & 0 & 0 & 0 \cr  0 & 0 & -9 & 3 & 0 & 0
& 0 & 0 & 0 & 0 \cr 0 & 0 & 0 & 0 & -1 & 3 & 0 & 0 & 0 & 0 \cr 0 &
0 & 0 & 0 & -3 & 17 & 0 & 0 & 0 & 0 \cr 0 & 0 & 0 & 0 & 0 & 0 & -1
& 3 & 0 & 0 \cr 0 & 0 & 0 & 0 & 0 & 0 & -3 & 17 & 0 & 0 \cr 0 & 0
& 0 & 0 & 0 & 0 & 0 & 0 & 3 & -9 \cr 0 & 0 & 0 & 0 & 0 & 0 & 0 & 0
& -9 & 3 \cr}\right)
\en
in the NDR scheme, and
\be
\label{rhv} \hat r_V^{\rm HV}=\left(\matrix{ {7\over 3} & -7 & 0 &
0 & 0 & 0 & 0 & 0 & 0 & 0 \cr -7 & {7\over 3} & 0 & 0 & 0 & 0 & 0
& 0 & 0 & 0 \cr 0 & 0 & {7\over 3} & -7 & 0 & 0 & 0 & 0 & 0 & 0
\cr 0 & 0 & -7 & {7\over 3} & 0 & 0 & 0 & 0 & 0 & 0 \cr 0 & 0 & 0
& 0 & -3 & 9 & 0 & 0 & 0 & 0 \cr 0 & 0 & 0 & 0 & 1 & {47\over 3} &
0 & 0 & 0 & 0 \cr 0 & 0 & 0 & 0 & 0 & 0 & -3 & 9 & 0 & 0 \cr 0 & 0
& 0 & 0 & 0 & 0 & 1 & {47\over 3} & 0 & 0 \cr 0 & 0 & 0 & 0 & 0 &
0 & 0 & 0 & {7\over 3} & -7 \cr 0 & 0 & 0 & 0 & 0 & 0 & 0 & 0 & -7
& {7\over 3} \cr}\right)
\en
in the HV scheme. Note that the 66 and 88 entries of $\hat r_V$
given in \cite{CCTY} are erroneous and have been corrected here.

\end{document}